\documentclass[sigconf,9pt]{acmart}
\usepackage{amsmath}

\usepackage{amsfonts}
\usepackage{pifont}
\usepackage{multirow}
\usepackage{enumitem}
\usepackage[table]{xcolor}
\usepackage[bottom]{footmisc} 
\setcopyright{acmlicensed}
\copyrightyear{2026}
\acmYear{2026}
\acmConference[DAC '26]{63rd ACM/IEEE Design Automation Conference}{July 26--29, 2026}{Long Beach, CA, USA}
\acmBooktitle{63rd ACM/IEEE Design Automation Conference (DAC '26), July 26--29, 2026, Long Beach, CA, USA}
\acmDOI{10.1145/3770743.3804061}
\acmISBN{979-8-4007-2254-7/2026/07}

\begin{document}



\title{LEAP: A Se\underline{l}f-Supervised P\underline{e}r-Cycle Toggle Prop\underline{a}gation Model \\ Supports Fast, Transferable, and Early Analysis of Layout \underline{P}ower}




\author{Wenkai Li$^\dagger$,\ Yuchao Wu$^\dagger$, Ziyan Guo, Yao Lu, Wenji Fang, Mengming Li, Zhiyao Xie}
\authornote{corresponding author, $^\dagger$equal contribution}
\affiliation{%
  \institution{Hong Kong University of Science and Technology}
  \country{\{wlidm, ywu092, zguoby, yludf, wfang838, mengming.li\}@connect.ust.hk, eezhiyao@ust.hk}
}


\begin{abstract}

Accurate power analysis is critical in VLSI design, as it directly impacts power optimization strategies. However, traditional approaches are often hindered by the substantial runtime required for per-cycle toggle propagation in the netlist, which propagates register toggle information through combinational logic. To address this, we propose LEAP, the first work to enable per-cycle toggle propagation prediction with both high accuracy and efficiency. This is achieved through a novel, linear-complexity graph transformer capable of simulating toggle propagation, along with specially designed self-supervised pre-training tasks that enable the model to capture circuit structure and functionality. LEAP achieves a 7.6$\times$ speedup over the EDA tool in toggle propagation, and attains a near-perfect area under the Precision-Recall curve (PR-AUC) of 0.99 for prediction results. 
Moreover, LEAP can be seamlessly integrated with other machine learning based power models into LEAP‑Power. This integration enables precise per‑cycle layout power prediction directly from post‑synthesis netlists, achieving a mean absolute percentage error (MAPE) of only 4.55\%. By bypassing toggle propagation in the netlist, LEAP‑Power delivers substantial runtime gains, running 5.3$\times$ faster than the model without LEAP.

\end{abstract}


\maketitle
\pagestyle{plain}
\section{INTRODUCTION}\label{sec:intro}

\begin{figure}[!t]
\centering
\includegraphics[width=0.48\textwidth]{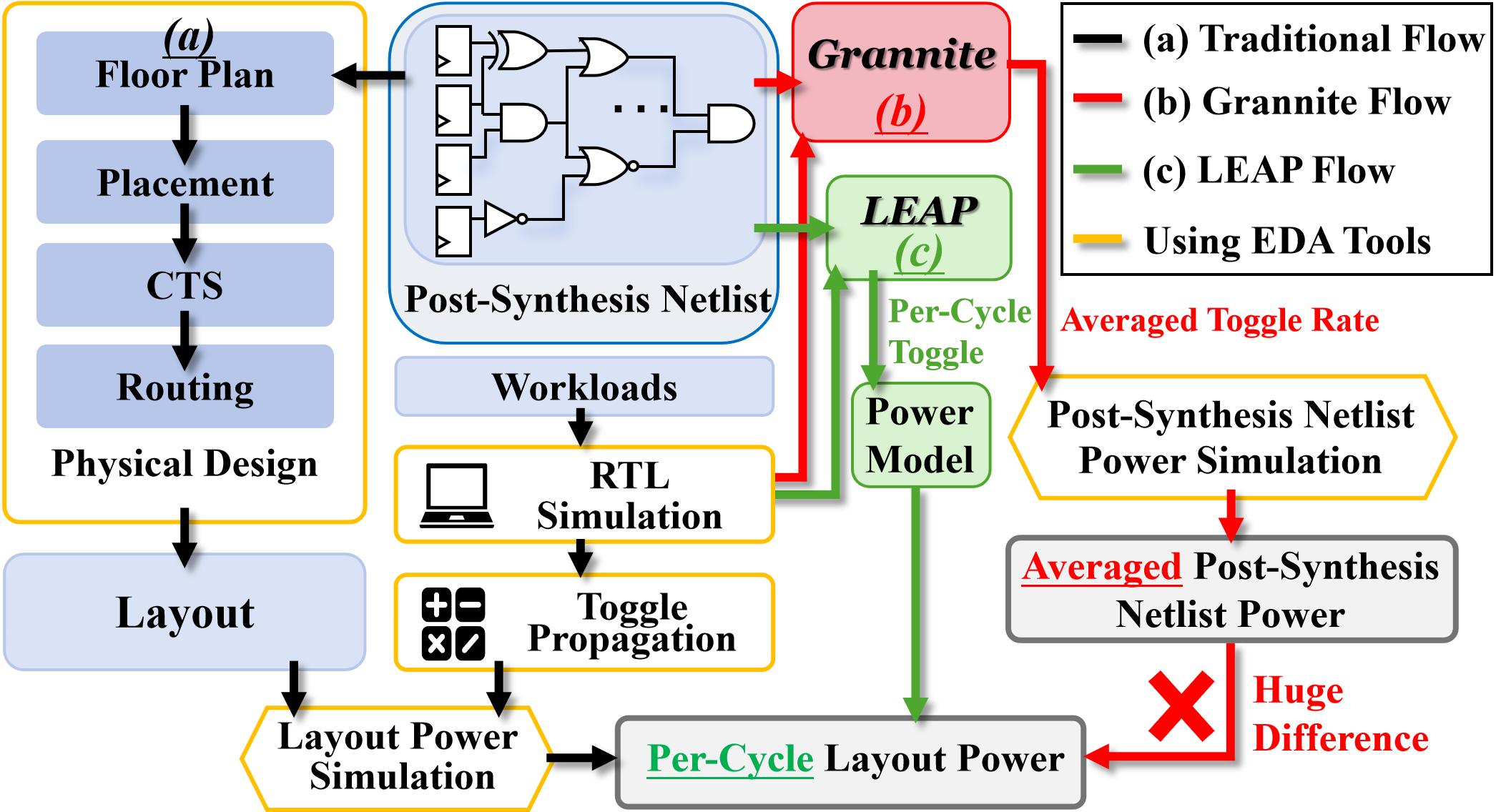}
\vspace{-.2in}
\caption{Overview of LEAP for per-cycle toggle propagation and cross-stage layout power prediction.
Traditional power estimation flow (a) for post-layout design is time-consuming due to physical design, per-cycle toggle propagation, and power simulation. Grannite~\cite{zhang2020grannite} (b) can only predict the averaged toggle rate and lacks the cross-stage power prediction ability.
LEAP (C) achieves significant acceleration over them.}
\vspace{-.1in}
\label{fig:flow}
\end{figure}

With the advancement of technology nodes, power has become an increasingly critical concern, as it directly affects chip voltage drop, energy consumption, and overall performance.  
Therefore, accurate power estimation is essential throughout the VLSI design flow. 
However, as shown in Fig.~\ref{fig:flow} (a), obtaining accurate per-cycle layout power with traditional EDA tools~\cite{vcs,Innovus,ptpx} is extremely time-consuming. The process first requires completing the full physical design flow to generate the layout. Next, toggle information must be produced: EDA tools perform RTL simulation to obtain register toggle information, followed by toggle propagation in the netlist to extend this information through the combinational logic. While RTL simulation is relatively fast, toggle propagation is significantly slower, and for operation system-level workloads, its runtime can even exceed that of physical design. Finally, once both the layout and toggle information are available, EDA tools conduct per-cycle power estimation, which is also computationally expensive. In summary, given post‑synthesis netlists and RTL simulation results, obtaining per-cycle layout power involves three time-consuming steps: \ding{182} physical design, \ding{183} per-cycle toggle propagation in the netlist, and \ding{184} per-cycle power estimation.

In recent years, numerous data-driven~\cite{du2024powpredi,fang2023masterrtl,fang2025nettag,ATLAS} power models have been proposed to bypass step \ding{182} and accelerate step \ding{184} via machine learning.
For works that support cross-design analysis, PowPredict~\cite{du2024powpredi}, MasterRTL~\cite{fang2023masterrtl}, and NetTAG~\cite{fang2025nettag}, predict averaged layout power across design stages.
ATLAS~\cite{ATLAS} further supports both cross-stage and per-cycle layout power prediction. 

Since step \ding{182} and \ding{184} are effectively bypassed or accelerated, \textbf{this makes step \ding{183} (per-cycle toggle propagation in the netlist) the current bottleneck}. There are a few works~\cite{zhang2020grannite,deepseq} to accelerate step \ding{183}: Grannite~\cite{zhang2020grannite} can predict the averaged toggle propagation (toggle rate) but cannot estimate per-cycle toggle propagation, as shown in Fig.~\ref{fig:flow} (b).
DeepSeq~\cite{deepseq} also predicts the averaged toggle rate; however, its applicability is restricted to netlists represented in the and-inverter graph (AIG) format. Since an AIG netlist contains only two types of gates—AND and inverter—this constraint significantly limits DeepSeq’s ability to capture toggle behavior in realistic gate-level netlists.
In summary, there is no existing work that can predict per-cycle toggle effectively.

\begin{figure*}[!t]
\vspace{-.7in}
\centering
\includegraphics[width=0.9\textwidth]{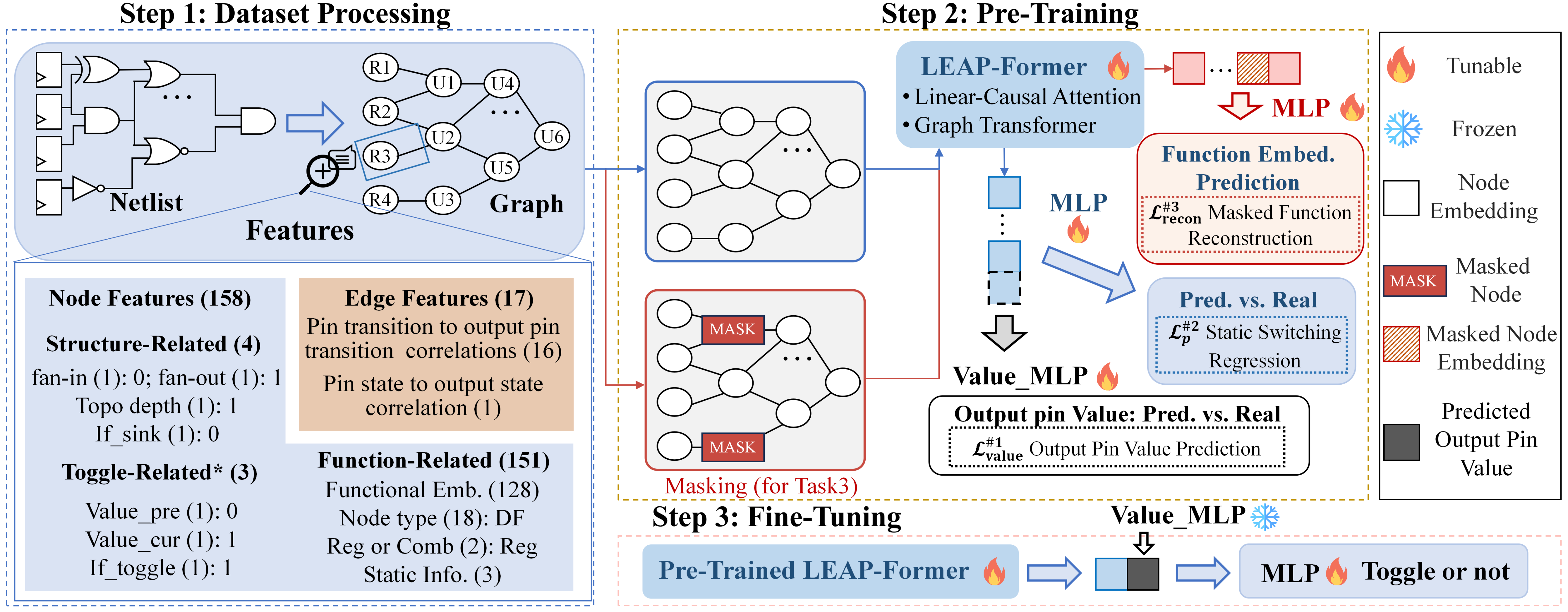} 
\vspace{-.1in}
\caption{LEAP Overview. LEAP includes three major steps: dataset processing (Section~\ref{sec:method-graphformat}), pre-training (Section~\ref{sec:method-pretrain}), and fine-tuning (Section~\ref{sec:method-finetune}). A detailed description of LEAP-Former is provided in Section~\ref{sec:method-LEAP-Former}. Only register nodes have toggle-related features, while these three features are set to zero for combinational nodes. 
}
\vspace{-.1in}
\label{fig:LEAP_overview}
\end{figure*}

In this work, we present LEAP, a vector-based per-cycle toggle propagation model built upon a linear-causal graph transformer. To the best of our knowledge, \textbf{LEAP is the first approach capable of accurately predicting per-cycle toggle propagation.}
Importantly, the ability of LEAP to learn and predict toggle propagation demonstrates that the model is effectively modeling the underlying functional and behavioral characteristics of gate-level circuits,  thereby enabling accurate and transferable predictions across different workloads and designs.
With the per-cycle toggle prediction results provided by LEAP, a wide range of downstream analyses become feasible, including per-cycle power estimation, dynamic timing analysis, and glitch power prediction.

LEAP not only provides an effective acceleration of Step \ding{183} (per-cycle toggle propagation in the netlist), but \textbf{also integrates seamlessly with other machine learning based power models}. When integrated, the entire flow from Step \ding{182} to Step \ding{184} can be significantly accelerated, with Step \ding{183} no longer constituting the primary bottleneck.
In essence, as shown in Fig.~\ref{fig:flow} (c), LEAP requires only the post-synthesis netlist and RTL simulation results to accurately predict per-cycle toggle propagation. Moreover, when integrated with other power models, the per-cycle layout power can be predicted with high accuracy.

LEAP introduces two major innovations:

     \textbf{1)} \texttt{\textbf{LEAP-Former:}}\
     We propose \textbf{LEAP-Former}, a graph transformer with linear complexity, specifically designed for directed graphs and to model the toggle propagation process. 
     The core is a novel \textbf{linear-causal attention} mechanism that mimics the actual toggle propagation: 
     We first define the topological depth of each node as the maximum distance from the node to its input register nodes. Each node, when computing its attention, can only see nodes with lower topological depths and cannot see those with higher topological depths. This aligns with the essence of directed graphs and toggle propagation, where each node can only access information from previous depths.
     A detailed comparison with other graph transformers and a full explanation of our attention mechanism are provided in Section~\ref{sec:method-LEAP-Former}.

     \textbf{2) Structure- and function-oriented pre-training \& Toggle propagation-oriented fine-tuning:} During the pre-training stage, LEAP employs the LEAP-Former as its backbone and utilizes a series of carefully designed self-supervised tasks, including functional prediction,  estimation of output pin value, etc.
     By leveraging large amounts of unlabeled design data, pre-training enables the model to learn both structural and functional properties of gate-level circuits in a data-efficient manner. This process not only provides an effective warm start for subsequent fine-tuning but also enhances the model’s generalization ability across different workloads and designs. As a result, LEAP can effectively capture intrinsic circuit behaviors and transfer its learned representations to diverse scenarios, thereby improving robustness and generalization of toggle propagation prediction. Building upon the pre-trained model, LEAP is further fine-tuned in a supervised manner, where the ground-truth labels are the toggle information (toggle or not) of each combinational gate. 
    

LEAP is evaluated on 12 out-of-order CPUs with gate counts ranging from 240,000 to 1,200,000, rather than on small-scale IP cores. The target workloads are also realistic, rather than hypothetical benchmarks, ensuring the robustness of the evaluation. LEAP achieves a 7.6$\times$ speedup over the EDA tool~\cite{ptpx} in toggle propagation, while attaining a near-perfect area under the Precision-Recall curve (PR-AUC) of 0.99 for toggle prediction results. Moreover, when integrated with other power models,  LEAP‑Power enables precise per‑cycle post‑layout power prediction directly from the post‑synthesis netlist, achieving a mean absolute percentage error (MAPE) of only 4.55\%. By bypassing toggle propagation, LEAP‑Power delivers dramatic runtime improvements, running 5.3$\times$ faster than the model without LEAP.

\pagestyle{empty}

\vspace{-.1in}
\section{PROBLEM FORMULATION AND OVERVIEW}\label{sec:method}

\subsection{Problem Formulation}

\noindent \textbf{Toggle Propagation Problem:} 
Given a post-synthesis netlist $P_n$ and all register toggle information $Tog_{\text{reg},t}$ after RTL simulation at cycle $t$, the objective is to predict the per-cycle combinational logic toggle information (whether toggle or not) in the post-synthesis netlist $P_n$. The model $f_{\text{toggle}}$  should simulate toggle propagation, and predict combinational logic toggle information at the cycle $t$:
\begin{equation}
f_{\text{toggle}}(P_n, Tog_{\text{reg},t}) \ \rightarrow\ {Tog}_{\text{comb},t}
\end{equation} 
where ${Tog}_{\text{comb},t}$ is the whole combinational logic toggle \emph{label} in post-synthesis netlist $P_n$ at the cycle $t$. 



\vspace{-.1in}
\subsection{Overview}\label{sec:method-overview}
Fig.~\ref{fig:LEAP_overview} provides an overview of LEAP to solve \textbf{Toggle Propagation Problem}. 
First, the entire netlist is transformed into a directed graph, where each gate is represented as a node, and the connections from output pins through nets to input pins are converted into graph edges. Each node in the constructed graph incorporates three types of features: functional, structural, and toggle. Those features are specifically tailored for LEAP-Former (Section~\ref{sec:method-graphformat}). 
Second, we present the internal mechanisms of LEAP-Former, explaining how it efficiently models toggle propagation with linear complexity (Section~\ref{sec:method-LEAP-Former}). 
Third, we introduce a structure‑ and function‑oriented pre‑training strategy (Section~\ref{sec:pretrain_finetune}) that comprehensively extracts functional and structural features from the directed graph, thereby maximizing the modeling capacity of LEAP‑Former. Building on this warm start of LEAP‑Former, we then present the fine‑tuning stage, which focuses specifically on toggle propagation.


\vspace{-.1in}
\section{DATASET PROCESSING}\label{sec:method-graphformat}

The objective of this step is to generate high-quality training data for the data-driven LEAP-Former model. We first transform the entire netlist into a directed graph, where each gate is represented as a node and connections from output pins through nets to input pins are modeled as graph edges. The direction of each edge follows the signal flow, originating from the output pin of a node at the preceding topological depth and terminating at the input pin of the connected node at the subsequent topological depth. 
A small number of gates, such as full adders, have multiple outputs; for these gates, we decompose them into individual nodes, with each node representing a single output and sharing the same input edges.

\textbf{Features in graph:}
Our graph incorporates both edge and node features. For edge features, we adopt a scheme similar to that used in Grannite~\cite{zhang2020grannite}, employing a 17-dimensional vector: one dimension encodes the correlation between the pin state and the output state, while the remaining 16 dimensions capture the correlations between pin transitions and output pin transitions.
As shown in Fig.~\ref{fig:LEAP_overview}, each node is annotated with the following three types of features:

     \textbf{1) Toggle-related features (3d (dimension)):}
These features are derived from RTL simulation results and are present only on register nodes; for combinational nodes, all three dimensions are set to zero. The three dimensions are: (i) the register node’s output pin value (0 or 1) at the previous cycle, (ii) its value at the current cycle, and (iii) whether a toggle occurred (0 or 1) at the current cycle. These features serve as the starting point for toggle propagation. 

     \textbf{2) Function-related features (151d):} 
\begin{itemize}[topsep=0pt, parsep=0pt]
    \item \textbf{Functional embedding (128d):} To enable the model to effectively capture the functionality of each gate, we convert its symbolic Boolean expression~\cite{fang2025nettag} into a functional embedding, which is then incorporated as part of the node features.
    For each gate, we extract the symbolic Boolean function of each gate from the standard cell library. For example, the output expression for the Z pin of AO22D4BWP is ((B1 B2)+(A1 A2)), and is represented as “AO22D4BWP Z ((B1 B2)+(A1 A2))”. We then employ an advanced general-purpose language embedding model (such as BGE-M3~\cite{bge-m3}) to convert this symbolic boolean expression in text into a 128d dense vector, termed the functional embedding. 

    \item \textbf{Static switching characteristics (3d):} Following Grannite \cite{zhang2020grannite}, we use 2d for the intrinsic state probabilities ($p_0$, $p_1$) and 1d for the intrinsic transition probability ($p_{01}$). 

    \item \textbf{Node type (20d):} We identify the 17 most common gate types in our dataset (ND, DF, AOI, etc.), and group all other types into an “Other” type. Each gate type is represented using an 18d one-hot code. To further enhance model understanding, we add a 2d one-hot code to encode whether a node is a register or a combinational gate.
    
\end{itemize}




     \textbf{3) Structure-related features (4d):}
These features capture each gate’s structure and topological depth: the node’s fan-in and fan-out counts; the node’s topological depth (the largest distance from the input registers); 
and whether the node is a sink node (final depth).
    

\begin{figure*}[!t]
\vspace{-.7in}
\centering
\includegraphics[width=0.88\textwidth]{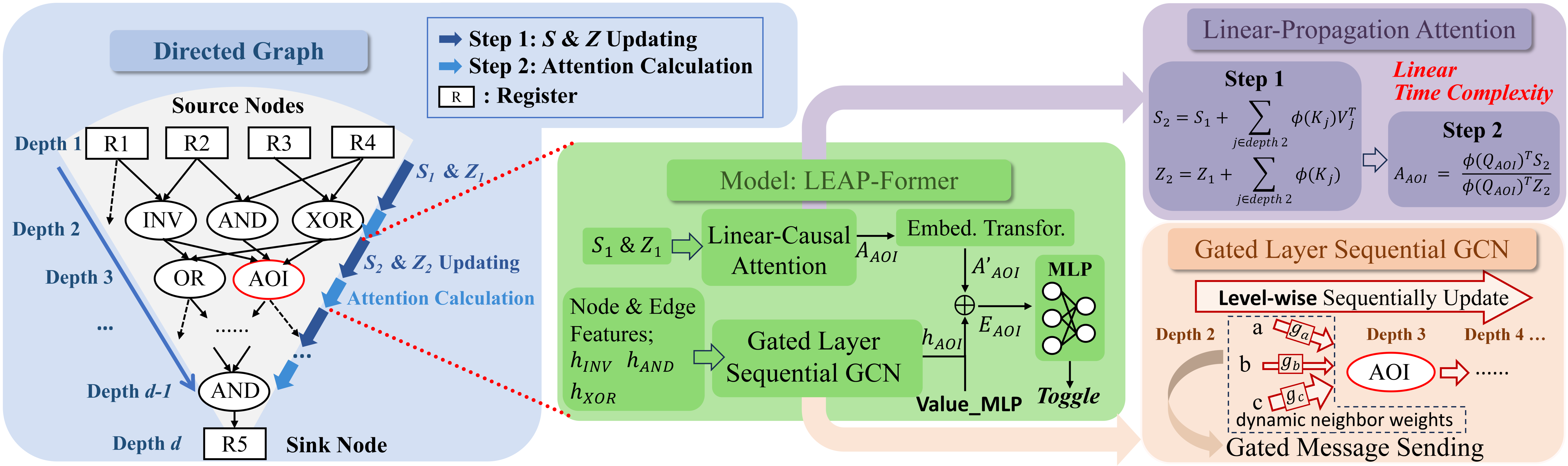} 
\vspace{-.1in}
\caption{LEAP-Former Overview. The core of LEAP‑Former is a linear‑causal attention mechanism designed for directed graphs to mimic toggle propagation. It operates in two main steps: 
$S$ and $Z$ updating, followed by attention calculation (Section~\ref{sec:4.2}). 
In addition, to convey structural information and thereby avoid the need for positional embeddings, a Gated GCN is integrated into the model (Section~\ref{sec:4.5}).
The combined node embedding 
$E$ is used to predict whether this node toggles at the current cycle.}
\vspace{-.1in}
\label{fig:LEAP-Former}
\end{figure*}

\vspace{-.1in}
\section{LEAP-FORMER FRAMEWORK}\label{sec:method-LEAP-Former}

In this section, we first discuss the limitations of existing graph transformers (Section~\ref{sec:4.1}), and then introduce the mechanism of our proposed graph transformer, \texttt{\textbf{LEAP-Former}} (Section~\ref{sec:4.2}). Finally, we explain how structural information is incorporated into LEAP-Former (Section~\ref{sec:4.5}).

\subsection{Existing Graph Transformers}\label{sec:4.1}
Graph Transformers (GTs) have recently emerged as powerful graph encoders. In contrast to Graph Neural Networks (GNNs), which are limited to message passing among local neighbors, GTs can capture long-range dependencies and implicit relationships. 

Currently, there are two main categories of GTs based on computational complexity: quadratic complexity and linear complexity.

\textbf{GTs with quadratic complexity:}
In this class~\cite{NodeFormer,wu2023difformer}, global all-pair attention is used: for each node, attention scores are computed between that node and every other node, and a weighted sum of other nodes' embeddings is used to update the node. This all-pair attention requires quadratic computation, resulting in significant computational bottlenecks and limiting its applicability to only small graphs (with up to hundreds of nodes).

\textbf{GTs with linear complexity:}
In this class~\cite{wu2023sgformer}, the mechanism can be viewed as introducing a virtual global node that connects to all nodes in the graph. Each node interacts only with this global node, rather than with every other node, and uses the result of this interaction to update its own embeddings.
This approach removes the need for global all-pair attention, resulting in linear complexity. However, under this attention mechanism, each node has access to information from the entire graph, which is not compatible with the directed graphs and the nature of toggle propagation. Because in toggle propagation, a gate can only receive information from previous topological depths, but cannot obtain information from subsequent depths.
\vspace{-.05in}
\subsection{Linear-causal Attention}\label{sec:4.2}

Our GT (LEAP-Former) adopts a linear-causal attention mechanism, which maintains linear complexity and is specifically designed for directed graphs with direction constraints as required by toggle propagation. 
We first compute the query $Q$, key $K$, and value $V$ for each node, and then explain the attention calculation in detail:

\textbf{QKV generation:} We first obtain the $Q$, $K$, and $V$ of every node. 
We map each node’s node feature and all its input edges' features into the same space using Multilayer Perceptrons (MLPs), and then sum them to obtain the node's edge-aware $Q$, $K$, and $V$.

\textbf{Linear-causal attention calculation:} 
Our model uses only a single attention layer with multi-head, rather than stacking multiple attention layers. The principle of the linear attention we adopt is from \cite{linear-attention}. In \cite{linear-attention}, for a \textbf{sequence} of length $N$ with a causal mask, the attention $A_i$ for token $i$ is calculated as follows:
\begin{equation}
    A_i = \frac{\phi(Q_i)^T \sum_{j=1}^{i} \phi(K_j) V_j^T}{\phi(Q_i)^T \sum_{j=1}^{i} \phi(K_j)} = \frac{\phi(Q_i)^T S_i}{\phi(Q_i)^T Z_i},\label{eq:linear}
\end{equation}
To simplify Eq.~\ref{eq:linear}, let $S_i = \sum_{j=1}^{i} \phi(K_j) V_j^T
$, and $Z_i = \sum_{j=1}^{i} \phi(K_j)
$. $\phi(x)=elu(x)+1$ is an activation function used to replace softmax~\cite{attention}, reducing quadratic complexity to linear complexity. $elu$ stands for exponential linear units~\cite{elu}. The causal mask ensures that node $i$ can only see the tokens before it and cannot see any tokens after it. 

After analyzing the linear attention formulation in Eq.~\ref{eq:linear} from \cite{linear-attention}, we observe that it can be naturally adapted to our task. Specifically, the entire sequence can be divided into segments, with each segment corresponding to all nodes at a particular topological depth.
Suppose node $i$ is at depth $d$; with the causal mask, node $i$ cannot see any nodes at deeper depths. We use $S_{d-1}$ and $Z_{d-1}$ to represent the accumulated $\phi(K_j) V_j^T$ and $\phi(K_j)$ of all nodes at depths before $d$. Thus, the calculation of $S_{d-1}$ and $Z_{d-1}$ at depth $d-1$ is:
\begin{align}
S_{d-1} &= \sum_{dp=1}^{d-1}\sum_{\text{j}\in dp} \phi(K_j) V_j^T, \quad
Z_{d-1} = \sum_{dp=1}^{d-1}\sum_{\text{j}\in dp}\phi(K_j) \label{eq:prefix}
\end{align}
where $dp$ is the depth position and $j$ is the node in depth $dp$.

Finally, the attention ${A}_i$ for node $i$ in depth $d$ is:
\begin{equation}
    A_i = \frac{\phi(Q_i)^T S_{d-1}}{\phi(Q_i)^T Z_{d-1}} \label{eq:linear_causal}
\end{equation}
where node $i$ can only see nodes from previous depths; it cannot see nodes at depth $d+1$ or deeper, nor can it see other nodes at the same depth. This aligns with the essence of directed graphs and toggle propagation, where each node can only access information from previous depths.

Fig.~\ref{fig:LEAP-Former} visualizes our attention operations: both the updates of $S$ and $Z$ and the attention computation are performed in order of increasing topological depth. For example, moving from depth $2$ to depth $3$ involves two steps. \textbf{Step 1, $S$ and $Z$ updating:} $S_2$ and $Z_2$ are derived from $S_1$ and $Z_1$ by simply adding the $\phi(K_j) V_j^T$ and $\phi(K_j)$, respectively, of all nodes at depth $2$. \textbf{Step 2, attention calculation:} for node \texttt{AOI} in depth $3$, we apply Eq.~\ref{eq:linear_causal} with $S_2$ and $Z_2$ to compute its attention $A_{AOI}$. The same operations are performed at depth $4$, and this process repeats until the final depth is reached.

With the attention output $A_{AOI}$ in Fig.~\ref{fig:LEAP-Former}, we follow the standard Transformer~\cite{attention} to obtain the final node embedding $A'_{AOI}$. Specifically, $A_{AOI}$ is processed through five consecutive operations: multi-head concatenation, output projection, residual connection, layer normalization, and a feed-forward network. 
In Fig.~\ref{fig:LEAP-Former}, we denote these five steps collectively as ``Embed. Transfor.”.

\subsection{Incorporation of Structural Information}\label{sec:4.5}

To convey structural information and thereby avoid the use of positional embeddings, we integrate a Gated Graph Convolutional Network (GCN)~\cite{ggnn,ggcn} into the model. The integration of attention and GCN in our method is inspired by SGFormer~\cite{wu2023sgformer}.
As shown in Fig.~\ref{fig:LEAP-Former}, the core role of this Gated GCN is dynamic neighbor weighting: each incoming message is scaled by a learned gate that depends on the source state, edge attributes, so neighbors no longer contribute equally but are adaptively attenuated or amplified. This gating mechanism naturally models toggle propagation, as not all driving signals influence the target gate; whether propagation occurs depends on gate type, input state, and edge connectivity. For example, suppose the function of an \texttt{AOI} gate is $Y = (ab+c)'$. Toggling $c$ alone can cause $Y$ to flip; then the gate $g_c$ associated with $c$ should have a higher weight.
The Gated GCN propagates information in a sequential (depth‑wise) manner; as a result, it can execute in parallel with the attention branch without incurring additional time. 
In brief, for \texttt{AOI}, the model takes as input its node features, edge features, and the embeddings ($h_{INV}$, $h_{AND}$, and $h_{XOR}$) from the preceding layer, and outputs the embedding $h_{AOI}$.
Finally, the embeddings ($h_{AOI}$ and $A'_{AOI}$) from the two branches are fused, either additively or via concatenation, and the combined embedding $E_{AOI}$ is used for the downstream toggle propagation task.

\begin{table*}[]
\vspace{-.7in}
\caption{The statistics of the gate counts for the twelve designs at the post-synthesis and post-layout stages.}
\vspace{-.1in}
\centering
\renewcommand{\arraystretch}{0.9}
\resizebox{.9\textwidth}{!}{
 \begin{tabular}{|c||c|c|c|c|c|c|c|c|c|c|c|c|c|}
\hline
                        & {\textbf{D1}} & {\textbf{D2}} & {\textbf{D3}} & {\textbf{D4}} & {\textbf{D5}} & {\textbf{D6}} & {\textbf{D7}} & {\textbf{D8}} & {\textbf{D9}} & {\textbf{D10}} & {\textbf{D11}} & {\textbf{D12}} \\ \hline \hline
\textbf{Post-synthesis} & 242012 & 267975 & 321116 & 399207 & 513305 & 570678 & 668339 & 838823 & 797875 & 1012021 & 1079086 & 1120456   \\ \hline
\textbf{Post-layout}  & 257969 & 285917 & 345872 & 418708 & 547915 & 605131 & 713740 & 929767 & 851145 & 1144901 & 1206049 & 1244124  \\ \hline
\end{tabular}}

\label{tbl:size}
\vspace{-.1in}
\end{table*}

\vspace{-.1in}

\section{LEAP PRE-TRAINING AND FINE-TUNING}\label{sec:pretrain_finetune}

\subsection{LEAP Pre-Training}\label{sec:method-pretrain}
This step is intended to equip LEAP-Former with essential knowledge of circuit structure and function prior to toggle propagation modeling in the fine-tuning stage.
Through the workload, we can obtain the per-cycle output pin value and the toggle label of each node. The goal of the fine-tuning stage is to predict the toggle.
We pre-train LEAP-Former using a multi-task self-supervised objective that explicitly teaches the encoder to capture two learning goals: \textbf{Goal 1} gate functionality, \textbf{Goal 2} structural and topological information. In Fig.~\ref{fig:LEAP_overview}, the pre-training stage consists of three tasks:

\textbf{Task} \ding{182} \textbf{Combinational node output pin value prediction (Goal 1):}  
We first let the model predict the per-cycle output pin value of each combinational node, which helps the model understand the function of different gates in each cycle. It is important to note that the value label (the value of combinational nodes) is only used during this task and will not appear in the fine-tuning stage. We use a ``Value\_MLP" (an MLP) to predict the value based on each node’s embedding via Mean Squared Error (MSE) loss. This task is layer-wise: the register output pin value for cycle $t$ is included in the register node features at the first depth, and is used to predict the output pin value of the next-depth combinational nodes; this process continues for subsequent depths. In the fine-tuning stage, we'll \textbf{freeze the ``Value\_MLP"} and use its predictions as an additional input feature to the MLP, together with the output embedding, to predict whether a node toggles. Through this task, the LEAP-Former is endowed with the ability to model and predict the functional behavior of each node. This task is denoted by $\mathcal{L}^{\#1}_{\text{value}}$.

\textbf{Task} \ding{183} \textbf{Static switching characteristic regression (Goal 1):} This regression task (MSE loss) requires predicting the static switching probabilities ($p_0$, $p_1$, $p_{01}$) of each node using the static features of its input edges. These probabilities reflect the intrinsic logical function of the gate and the statistical relationships between its inputs and outputs. By performing this task, the model is encouraged to capture both the functional semantics and the static switching characteristics of each gate, thereby enhancing its ability to model gate function. This task is denoted by $\mathcal{L}^{\#2}_{p}$.

\textbf{Task} \textbf{\ding{184} Masked functional embedding reconstruction (Goal 1 \& Goal 2):}  
This task randomly masks the 128-dimensional functional embedding of nodes and requires the model to reconstruct it, thereby forcing the model to infer each node’s functional representation from its structural neighborhood and context via MSE loss. This encourages the model to simultaneously capture both the functional semantics of nodes and their structural context, thereby enhancing generalization capability. This task is denoted by $\mathcal{L}^{\#3}_{\text{recon}}$.

Finally, we formulate the overall pre-training objective of our model by jointly optimizing across the three tasks with specific coefficients $\alpha$:
\begin{equation}
\mathcal{L} \;=\; \alpha_{\text{value}}\mathcal{L}^{\#1}_{\text{value}}
+ \alpha_{p}\mathcal{L}^{\#2}_{p}
+ \alpha_{\text{recon}}\mathcal{L}^{\#3}_{\text{recon}}
\end{equation}

Together, these tasks constitute a unified pre‑training framework that equips LEAP‑Former to reason over both gate functionality and structural hierarchy. This rich initialization lays the foundation for accurate toggle propagation modeling in downstream fine-tuning.


\vspace{-.10in}
\subsection{LEAP Fine-Tuning}\label{sec:method-finetune}

Building on the pre-trained LEAP-Former backbone, the fine-tuning stage follows a strict toggle propagation paradigm, where each combinational node receives as input the information propagated from the previous depths, along with its own node and edge features. These inputs are processed by the pre-trained LEAP-Former to generate the embedding for each node. Throughout this process, the three toggle-related features of the first-depth register nodes are iteratively passed downstream, serving as the initial conditions for toggle propagation. Meanwhile, the frozen ``Value\_MLP'' (in Fig.~\ref{fig:LEAP_overview} and Fig.~\ref{fig:LEAP-Former}) predicts the output pin value for each combinational node. Subsequently, each combinational node’s embedding and its output pin value are fed into a simple MLP, which performs a binary classification to determine whether a toggle occurs at that cycle.                                                                                                                                                       

\pagestyle{empty}
\vspace{-.10in}
\section{EXPERIMENTAL RESULTS} \label{sec:expr}

\subsection{Experiment Setup}


We utilized the TSMC 40nm standard cell library~\cite{URL:tmsc40nm}. The logic synthesis is executed at a clock frequency of 1GHz using Design Compiler\textsuperscript{\textregistered}\cite{design-compilier}, followed by 
physical design using Innovus\textsuperscript{\textregistered}~\cite{Innovus}. 
RTL simulation on workloads is based on  VCS\textsuperscript{\textregistered}\cite{vcs}.
The ground-truth per-cycle toggle propagation and layout power simulation are performed using Synopsys PrimeTime PX\textsuperscript{\textregistered}~\cite{ptpx} (PTPX\textsuperscript{\textregistered}). 

Our dataset consists of 12 realistic out-of-order CPU designs. Table~\ref{tbl:size} presents the cell count statistics for 12 different designs at both the post-synthesis and post-layout stages. 
For workloads, we extract 300-cycle segments from the following benchmarks: the Median Filter benchmark (Median), the Towers of Hanoi benchmark (Towers), and the Vector-Vector Add benchmark (Vvadd).
We reproduced the Grannite~\cite{zhang2020grannite} model framework as our baseline. It is important to note that original Grannite does not support per-cycle toggle propagation and is limited to average toggle rate prediction. Furthermore, Grannite’s original feature set does not include functionality- or structure-related features. For a fair comparison, we provided Grannite with the full set of features from our dataset. 
Our pre-training and fine-tuning are conducted on a Linux machine equipped with 8 NVIDIA RTX 4090D GPUs. All evaluations are performed on a single GPU.

\textbf{Evaluation Metrics.} For our toggle propagation (binary classification) task, we adopt two standard evaluation metrics: PR-AUC~\cite{mao2019ap} and ROC-AUC~\cite{liang2020routing}. PR-AUC is the area under the Precision-Recall (PR) curve. ROC-AUC is the area under the Receiver Operating Characteristic (ROC) curve.  
Both metrics are upper-bounded by 1, and higher values indicate better performance.
Precision measures the proportion of correctly predicted positive samples among all predicted positives, while Recall measures the proportion of correctly predicted positives among all actual positives. 
For this binary classification task, \textbf{improvements in PR-AUC are significantly more meaningful than those in ROC-AUC}~\cite{roc_pr}. Because, in any given cycle, toggling combinational nodes account for only 3\%–10\% of all nodes, resulting in a highly imbalanced distribution between positive and negative samples. Under such conditions, ROC-AUC may remain relatively high in accuracy even when the model fails to capture rare positive cases, whereas PR-AUC directly reflects the model’s ability to identify these sparse toggling events. Therefore, PR-AUC serves as a more sensitive and relevant metric for evaluating per-cycle toggle prediction performance.



\definecolor{lightgreen}{RGB}{200, 245, 180}

\begin{table}[t]
\caption{PR-AUC \& ROC-AUC results under three workloads. PR here refers to PR-AUC. ROC here refers to ROC-AUC.}
\vspace{-.1in}
\renewcommand{\arraystretch}{0.95}
\centering
\arrayrulecolor{black}
\resizebox{.48\textwidth}{!}{
\setlength{\tabcolsep}{1.5pt}
\begin{tabular}{|c|cc|cc|cc|cc|cc|cc|}

\hline
\multirow{3}{*}{\textbf{Design}} 
    & \multicolumn{6}{c|}{\textbf{Grannite~\cite{zhang2020grannite}}} & \multicolumn{6}{c|}{\textbf{LEAP}}
 \\ \cline{2-13} 
    & \multicolumn{2}{c|}{\textbf{Median}} & \multicolumn{2}{c|}{\textbf{Towers}} & \multicolumn{2}{c|}{\textbf{Vvadd}} 
    & \multicolumn{2}{c|}{\textbf{Median}} & 
\multicolumn{2}{c|}{\textbf{Towers}} & 
\multicolumn{2}{c|}{\textbf{Vvadd}}
\\ \cline{2-13}
    & \textbf{PR} & \textbf{ROC} & \textbf{PR} & \textbf{ROC} & \textbf{PR} & \textbf{ROC} 
    &  \textbf{PR} &  \textbf{ROC} &  \textbf{PR} &  \textbf{ROC} &  \textbf{PR} &  \textbf{ROC} \\ \hline\hline
D1  & 0.83 & 0.97 & 0.83 & 0.97 & 0.80 & 0.97 &  0.99 &  1.00 &  0.99 &  1.00 &  0.99 &  1.00 \\ \hline
D2  & 0.84 & 0.97 & 0.83 & 0.98 & 0.82 & 0.98 &  0.99 &  1.00 &  1.00 &  1.00 &  1.00 &  1.00 \\ \hline
D3  & 0.80 & 0.97 & 0.82 & 0.97 & 0.83 & 0.97 &  0.99 &  1.00 &  0.99 &  1.00 &  0.99 &  1.00 \\ \hline
D4  & 0.81 & 0.96 & 0.83 & 0.98 & 0.83 & 0.98 &  0.99 &  1.00 &  1.00 &  1.00 &  1.00 &  1.00 \\ \hline
D5  & 0.83 & 0.97 & 0.82 & 0.97 & 0.84 & 0.97 &  0.99 &  1.00 &  0.99 &  1.00 &  0.99 &  1.00 \\ \hline
D6  & 0.85 & 0.98 & 0.87 & 0.98 & 0.85 & 0.98 &  0.99 &  1.00 &  1.00 &  1.00 &  0.99 &  1.00 \\ \hline
D7  & 0.84 & 0.97 & 0.85 & 0.98 & 0.83 & 0.97 &  0.99 &  1.00 &  1.00 &  1.00 &  0.99 &  1.00 \\ \hline
D8  & 0.80 & 0.96 & 0.84 & 0.97 & 0.84 & 0.97 &  1.00 &  1.00 &  0.99 &  1.00 &  0.99 &  1.00 \\ \hline
D9  & 0.83 & 0.98 & 0.83 & 0.98 & 0.84 & 0.97 &  1.00 &  1.00 &  1.00 &  1.00 &  0.99 &  1.00 \\ \hline
D10 & 0.82 & 0.97 & 0.84 & 0.97 & 0.84 & 0.98 &  0.99 &  1.00 &  0.99 &  1.00 &  0.99 &  1.00 \\ \hline
D11 & 0.83 & 0.97 & 0.85 & 0.98 & 0.82 & 0.98 &  0.99 &  1.00 &  0.99 &  1.00 &  0.99 &  1.00 \\ \hline
D12 & 0.81 & 0.97 & 0.84 & 0.98 & 0.84 & 0.98 &  1.00 &  1.00 &  0.99 &  1.00 &  0.99 &  1.00 \\ \hline\hline
\textbf{AVG} & \textbf{0.82} & \textbf{0.97} & \textbf{0.84} & \textbf{0.98} & \textbf{0.83} & \textbf{0.97} 
&  \cellcolor{lightgreen}\textbf{0.99} &  \cellcolor{lightgreen}\textbf{1.00} &  \cellcolor{lightgreen}\textbf{0.99} &  \cellcolor{lightgreen}\textbf{1.00} &  \cellcolor{lightgreen}\textbf{0.99} &  \cellcolor{lightgreen}\textbf{1.00} \\ \hline
\end{tabular}
}
\vspace{-.2in}
\label{tbl:ap_results}
\end{table}

\vspace{-.10in}
\subsection{Toggle Prediction Results}\label{sec:toggle_result}

Table~\ref{tbl:ap_results} presents the detailed prediction results of LEAP and Grannite, where both PR-AUC and ROC-AUC are averaged over 300 cycles. We employ a 2‑fold validation scheme, where designs are partitioned into two groups: In Setting 1, the training set includes designs D2, 4, 7, 9, and 11 with the Median workload, while the test set covers all remaining designs and three workloads. In Setting 2, designs D1, 3, 5, 6, 8, 10, and 12 with the Median workload are used for training, and other designs with all three workloads are used for testing.

\textbf{High-accuracy, cross-design, and cross-workload capability of LEAP:} As shown in the Table~\ref{tbl:ap_results}, LEAP consistently achieves highly accurate predictions across all three workloads, with near-perfect averaged PR-AUC scores reaching 0.99. Compared to Gran-nite’s PR-AUC range of 0.82–0.84, this represents a substantial improvement, demonstrating LEAP’s superior ability to model toggle propagation. Furthermore, the 2-fold results confirm LEAP’s strong cross-design generalization. Even when trained solely on the Median workload, LEAP maintains a PR-AUC of 0.99 on the other two workloads, highlighting its robust cross-workload capability. Considering the class imbalance, both LEAP and Grannite perform well in terms of ROC-AUC. 
However, LEAP achieves a near‑perfect ROC-AUC of 0.999, while Grannite remains in the 0.97–0.98 range.

\vspace{-.1in}
\subsection{Runtime Comparison}

\begin{table}[]
\vspace{-.7in}
\caption{Toggle propagation runtime comparison.}
\vspace{-.1in}
\renewcommand{\arraystretch}{1.0}
\centering
\resizebox{.48\textwidth}{!}{
\begin{tabular}{|c|ccccccc|c|}
\hline
\multirow{2}{*}{\textbf{Method}} & \multicolumn{8}{c|}{\textbf{Runtime (Unit: Second)}} \\ \cline{2-9}
  & \textbf{D1} & \textbf{D3} & \textbf{D5} & \textbf{D6} & \textbf{D8} & \textbf{D10} & \textbf{D12} & \textbf{AVG} \\ \hline\hline
PTPX~\cite{ptpx}     & 61.3 & 70.8 & 85.0 & 105.4  & 122.5 & 161.0 & 164.6 & 110.1 \\ \hline
Grannite~\cite{zhang2020grannite} & 6.8  & 10.0 & 13.7 & 10.4  & 12.6  & 11.8  & 17.1  & 11.8  \\ \hline
 LEAP    & 8.6  & 11.8 & 14.3 & 15.4  & 15.4  & 17.1  & 19.0  & 14.5  \\ \hline
\end{tabular}
}
\vspace{-.10in}
\label{tbl:runtime_comparison}
\end{table}

\begin{table}[b]
\vspace{-.1in}
\caption{Power prediction errors. LEAP-Power incurs only a negligible loss in power prediction accuracy while delivering a substantial speedup, as shown in Table~\ref{tbl:runtime_30000cycle}. }
\vspace{-.1in}
\renewcommand{\arraystretch}{0.95}
\centering
\resizebox{.48\textwidth}{!}{
\begin{tabular}{|c|cccc|cccc|}
\hline
\multirow{3}{*}{\textbf{Design}} & \multicolumn{8}{c|}{\textbf{Average MAPE (\%) of Three Workloads}} \\ \cline{2-9} 
 & \multicolumn{4}{c|}{\textbf{ATLAS~\cite{ATLAS}}} & \multicolumn{4}{c|}{\textbf{LEAP-Power}} \\ \cline{2-9} 
 & \textbf{Reg} & \textbf{CLK} & \textbf{Comb} & \textbf{Total} & \textbf{Reg} & \textbf{CLK} & \textbf{Comb} & \textbf{Total} \\ \hline\hline
D1   & 4.08 & 10.32 & 9.58  & 5.78 & 4.08 & 10.32 & 10.27 & 5.98  \\ \hline
D3   & 5.02 & 12.63 & 7.19  & 5.20 & 5.02 & 12.63 & 8.04  & 5.46  \\ \hline
D5   & 4.73 & 5.39  & 7.26  & 4.71 & 4.73 & 5.39  & 7.68  & 5.02  \\ \hline
D6   & 3.00 & 4.74  & 8.03  & 3.75 & 3.00 & 4.74  & 8.89  & 3.84  \\ \hline
D8   & 3.55 & 1.47  & 10.42 & 4.09 & 3.55 & 1.47  & 11.02 & 4.65  \\ \hline
D10  & 3.39 & 5.69  & 5.72  & 2.49 & 3.39 & 5.69  & 6.41  & 2.69  \\ \hline
D12  & 2.10 & 6.26  & 9.06  & 3.82 & 2.10 & 6.26  & 9.55  & 4.22  \\ \hline\hline
\textbf{AVG} & \textbf{3.70} & \textbf{6.64} & \textbf{8.32} & \textbf{4.26} & \textbf{3.70} & \textbf{6.64} & \textbf{8.84} & \textbf{4.55} \\ \hline
\end{tabular}
}
\label{tbl:mape_error}
\end{table}

Table~\ref{tbl:runtime_comparison} compares the runtime of toggle propagation in Setting 1\footnote{Due to space limitations, we only present the results of Setting 1 here, while those of Setting 2 show similar trends. The same applies to Table~\ref{tbl:mape_error} and Table~\ref{tbl:runtime_30000cycle}.} across three methods: PTPX~\cite{ptpx}, Grannite~\cite{zhang2020grannite}, and LEAP. PTPX is a commercial EDA tool used for toggle propagation and serves as the source of our golden labels. The reported runtime corresponds to the average time required by the three methods to perform toggle propagation over three 300-cycle workloads.
Compared to PTPX, LEAP achieves a 7.6$\times$ speedup, and while its acceleration over Grannite (9.3$\times$) is slightly lower, this is expected and justified. LEAP incorporates a more sophisticated yet still linear-complexity LEAP-Former, which trades minimal runtime overhead for substantial gains in prediction accuracy. 

\vspace{-.1in}
\subsection{Power Prediction Results}

LEAP can be seamlessly integrated with other power models. We use LEAP-Power, which integrates LEAP with ATLAS~\cite{ATLAS}, to predict per-cycle layout power starting from post-synthesis netlists. LEAP-Power achieves a substantial runtime speedup while incurring only a negligible loss in power prediction accuracy. We use Setting 1 for training and testing, and Table~\ref{tbl:mape_error} shows the average prediction results across three workloads, where we use MAPE to report errors. In ATLAS~\cite{ATLAS}, the total power is decomposed into three components: register, clock tree (CLK), and combinational logic. Since toggle propagation only affects combinational logic, ATLAS and LEAP-Power produce the same prediction results for registers and the clock tree. After integrating LEAP, the MAPE for combinational logic and total power decreases by just 0.52\% and 0.29\%, respectively. This demonstrates that LEAP can work perfectly with other power models. Moreover, LEAP-Power can skip the toggle propagation process entirely, greatly accelerating prediction time. Table~\ref{tbl:runtime_30000cycle} reports the runtime comparison across methods. 
For a 30,000‑cycle workload, compared with the traditional flow in Fig.~\ref{fig:flow}(a), ATLAS achieves a speedup of 12$\times$, while LEAP‑Power delivers a speedup of 67$\times$ by bypassing toggle propagation, making it 5.3$\times$ faster than ATLAS under this setting.

\begin{table}[t]
\vspace{-.7in}
\caption{Power prediction runtime comparison}
\vspace{-.1in}
\renewcommand{\arraystretch}{1.0}
\centering
\resizebox{.48\textwidth}{!}{
\begin{tabular}{|c|ccccccc|c|}
\hline
\multirow{2}{*}{\textbf{Method}} & \multicolumn{8}{c|}{\textbf{Runtime for 30,000 Cycles (Unit: Minute)}} \\ \cline{2-9}
& \textbf{D1} & \textbf{D3} & \textbf{D5} & \textbf{D6} & \textbf{D8} & \textbf{D10} & \textbf{D12} & \textbf{AVG} \\ \hline\hline
Traditional Flow   & 995 & 1392  & 2133 & 2382 & 2927 & 3451 & 3516 & 2399 \\ \hline
ATLAS~\cite{ATLAS}       & 106  & 125   & 151   & 180  & 214  & 282  & 289  & 192  \\ \hline
\rowcolor{lightgreen} LEAP-Power  & 20  & 26   & 33   & 36   & 40   & 45   & 49   & 36   \\ \hline
\end{tabular}
}
\vspace{-.1in}
\label{tbl:runtime_30000cycle}
\end{table}


\begin{figure}[b]
\centering
\includegraphics[width=0.48\textwidth]{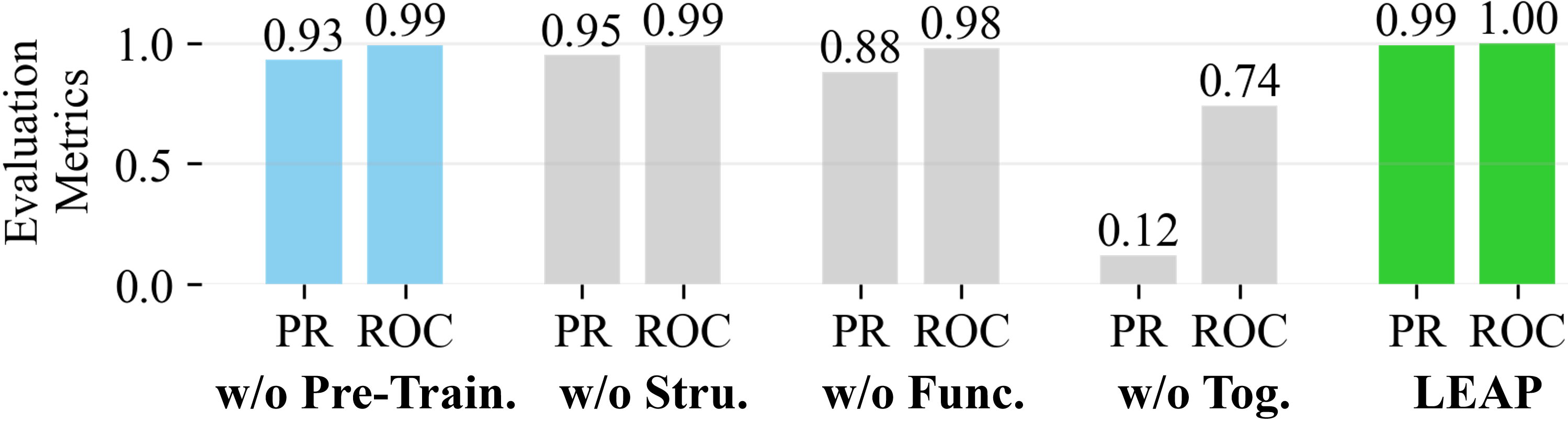}
\vspace{-.2in}
\caption{Ablation study on pre-training and feature types.}

\label{fig:ablation_feature}
\end{figure}

\subsection{Ablation Study}

We conduct two ablation studies to investigate the source of LEAP’s effectiveness. 
These experiments follow the same 2-fold validation setup as described in Section~\ref{sec:toggle_result}. The results represent the average PR-AUC and ROC-AUC across 12 designs and 3 workloads in Fig.~\ref{fig:ablation_feature}.

\textbf{Ablation Study on Pre-training stage:} In the left blue columns of Fig.~\ref{fig:ablation_feature}, we remove the entire pre-training stage and exclude the Value\_MLP used during fine-tuning. Instead, we apply supervised toggle propagation prediction directly using LEAP-Former. Compared to the full LEAP model with pre-training, averaged per-cycle PR-AUC scores drop from 0.99 to 0.93 across all three workloads. This clearly demonstrates that LEAP benefits significantly from pre-training, which enables the model to learn gate-level functionality and netlist structure prior to fine-tuning.

\textbf{Ablation Study on Three Feature Types:} In the grey columns of Fig.~\ref{fig:ablation_feature}, we individually zero out each of the three feature types in the dataset and report the averaged per-cycle PR-AUC across 12 designs and three workloads. Since toggle-related features are only present on register nodes (combinational nodes' toggle features are all zero), we zero out toggle-related features exclusively on registers. When all toggle-related features on register nodes are removed, PR-AUC drops dramatically from 0.99 to 0.12—effectively rendering predictions inaccurate. This highlights that LEAP learns to propagate toggle-related features layer by layer from registers through the combinational logic, successfully and accurately modeling the toggle propagation process.
We also observe that removing structure-related and function-related features leads to PR-AUC drops of 0.04 and 0.11, respectively. This confirms the effectiveness of both feature types, with function-related features playing a more critical role in helping the model understand toggle propagation.

\pagestyle{empty}

\vspace{-.1in}
\section{Conclusion}\label{sec:concl}
We have proposed LEAP, the first framework to enable per‑cycle toggle propagation prediction, achieving a near-perfect PR-AUC of 0.99 and a 7.6$\times$ speedup over the commercial EDA tool. LEAP attains high accuracy, cross‑design generalization, and cross‑workload capability through a novel linear‑complexity graph transformer that simulates toggle propagation, further enhanced by self‑supervised pre‑training that captures circuit structure and functionality prior to fine‑tuning.
Moreover, LEAP can be seamlessly integrated with other power models, ensuring high accuracy while delivering significant speedup by bypassing toggle propagation.


\pagestyle{empty}
\section{ACKNOWLEDGEMENT}

This work is supported by National Natural Science Foundation of China (NSFC) 62304192, Hong Kong Research Grants Council (RGC) CRF-YCRG C6003-24Y, and T46-415/25-R. It was partially conducted by ACCESS – AI Chip Center for Emerging Smart Systems, supported by the InnoHK initiative of the Innovation and Technology Commission of the Hong Kong Special Administrative Region Government.

\newpage
\bibliographystyle{ACM-Reference-Format}
\bibliography{ref}

@inproceedings{ATLAS,
author = {Wenkai Li and Yao Lu and Wenji Fang and Jing Wang and Qijun Zhang and Zhiyao Xie},
title = {ATLAS: A Self-Supervised and Cross-Stage Netlist Power
Model for Fine-Grained Time-Based Layout Power Analysis},
year = {2025},
booktitle = {Proceedings of the 62th ACM/IEEE Design Automation Conference},
}

@inproceedings{roc_pr,
author = {Davis, Jesse and Goadrich, Mark},
title = {The relationship between Precision-Recall and ROC curves},
year = {2006},
booktitle = {Proceedings of the 23rd International Conference on Machine Learning}
}

@inproceedings{linear-attention,
author = {Katharopoulos, Angelos and Vyas, Apoorv and Pappas, Nikolaos and Fleuret, Fran\c{c}ois},
title = {Transformers are RNNs: fast autoregressive transformers with linear attention},
year = {2020},
booktitle = {Proceedings of the 37th International Conference on Machine Learning}
}

@inproceedings{attention,
author = {Vaswani, Ashish and Shazeer, Noam and Parmar, Niki and Uszkoreit, Jakob and Jones, Llion and Gomez, Aidan N. and Kaiser, \L{}ukasz and Polosukhin, Illia},
title = {Attention is all you need},
year = {2017},
booktitle = {Proceedings of the 31st International Conference on Neural Information Processing Systems}
}

@article{elu,
  title={Fast and accurate deep network learning by exponential linear units (elus)},
  author={Clevert, Djork-Arn{\'e} and Unterthiner, Thomas and Hochreiter, Sepp},
  journal={arXiv:1511.07289},
  year={2015}
}

@misc{ggcn,
      title={Residual Gated Graph ConvNets}, 
      author={Xavier Bresson and Thomas Laurent},
      year={2018}, 
}

@inproceedings{ggnn,
  title={Gated Graph Sequence Neural Networks},
  author={Li, Yujia and Zemel, Richard and Brockschmidt, Marc and Tarlow, Daniel},
  booktitle={Proceedings of ICLR'16},
  year={2016}
}

@inproceedings{mao2019ap,
  title={A delay metric for video object detection: What average precision fails to tell},
  author={Mao, Huizi and Yang, Xiaodong and Dally, William J},
  booktitle={Proceedings of the IEEE/CVF International Conference on Computer Vision},
  year={2019}
}

@misc{bge-m3,
      title={BGE M3-Embedding: Multi-Lingual, Multi-Functionality, Multi-Granularity Text Embeddings Through Self-Knowledge Distillation}, 
      author={Jianlv Chen and Shitao Xiao and Peitian Zhang and Kun Luo and Defu Lian and Zheng Liu},
      journal={arkiv:2402.03216},
      year={2024}
}

@inproceedings{wu2023sgformer,
    title={SGFormer: Simplifying and Empowering Transformers for Large-Graph Representations},
    author={Qitian Wu and others},
    booktitle={Advances in Neural Information Processing Systems (NeurIPS)},
    year={2023}
}

@inproceedings{
wu2023difformer,
title={{DIFF}ormer: Scalable (Graph) Transformers Induced by Energy Constrained Diffusion},
author={Qitian Wu and others},
booktitle={The Eleventh International Conference on Learning Representations },
year={2023}
}

@inproceedings{NodeFormer,
author = {Wu, Qitian and Zhao, Wentao and others},
title = {NodeFormer: a scalable graph structure learning transformer for node classification},
year = {2024},
booktitle = {Proceedings of the 36th International Conference on Neural Information Processing Systems}

}

@inproceedings{fang2023masterrtl,
  title={MasterRTL: A Pre-Synthesis PPA Estimation Framework
for Any RTL Design},
  author={Fang, Wenji and others},
  booktitle={Proc. IEEE/ACM Int. Conf. Comput. Aided Design (ICCAD),},
  year={2023},
pages={1-9},
}

@INPROCEEDINGS{deepseq,
  author={Khan, Sadaf and Shi, Zhengyuan and Li, Min and Xu, Qiang},
  booktitle={2024 Design, Automation \& Test in Europe Conference \& Exhibition (DATE)}, 
  title={DeepSeq: Deep Sequential Circuit Learning}, 
  year={2024}
}

@misc{design-compilier,
 title = {{Design Compiler® RTL Synthesis}},
 howpublished = "\nolinkurl{https://www.synopsys.com/implementation-and-signoff/rtl-synthesis-test/design-compiler-nxt.html}",
    year = "2021",
 }

@inproceedings{fang2025nettag,
  title={NetTAG: A Multimodal RTL-and-Layout-Aligned Netlist Foundation Model via Text-Attributed Graph},
  author={Fang, Wenji and Li, Wenkai and Liu, Shang and Lu, Yao and Zhang, Hongce and Xie, Zhiyao},
  booktitle={Proceedings of 2025 IEEE/ACM Design Automation Conference (DAC)},
  year={2025}
}

@misc{vcs,
 title = {{VCS® functional verification solution}},
 howpublished = "\nolinkurl{https://www.synopsys.com/verification/simulation/vcs.html}",
    year = "2021",
 }

@inproceedings{du2024powpredi,
  title={PowPrediCT: Cross-Stage Power Prediction with Circuit-Transformation-Aware Learning},
  author={Du, Yufan and others},
  booktitle={Proc. Design Automation Conf. (DAC)},
  year={2024},
pages={1-6},
}

@inproceedings{zhang2020grannite,
  title={GRANNITE: Graph neural network inference for transferable power estimation},
  author={Zhang, Yanqing and others},
  booktitle={Proc. Design Automation Conf. (DAC)},
  year={2020}
}

@misc{ptpx,
  author = {Synopsys},
  title = {{PrimePower: RTL} to Signoff Power Analysis},
  url = {https://www.synopsys.com/implementation-and-signoff/signoff/primepower.html},
  year = {2023},
}

@manual{URL:tmsc40nm,
  title   = "TSMC 40nm LP process technology",
  address = "https://www.tsmc.com/english/dedic atedFoundry/technology/logic/l\_40nm",
  year = {2008}
}

@misc{Innovus,
  author = {Cadence},
  title = {Innovus Implementation System},
  year = {2021},
}

@inproceedings{liang2020routing,
  title={Routing-free crosstalk prediction},
  author={Liang, Rongjian and Xie, Zhiyao and Jung, Jinwook and Chauha, Vishnavi and Chen, Yiran and Hu, Jiang and Xiang, Hua and Nam, Gi-Joon},
  booktitle={International Conference on Computer Aided Design (ICCAD)},
  year={2020},
}

\end{document}